\documentclass[a4paper, 10pt, conference]{ieeeconf}      

\usepackage{graphics} 
\usepackage{epsfig} 
\usepackage{mathptmx} 
\usepackage{times} 
\usepackage{amsmath} 
\usepackage{amssymb}  
\usepackage[justification=centering]{caption}
\title{\LARGE \bf
On Physically Secure and Stable Slotted ALOHA System }

\author{Yunus Sarikaya, Ozgur
Ercetin \\
Faculty of Engineering and Natural Sciences Sabanci University,
Istanbul, Turkey}

\def\QED{\mbox{\rule[0pt]{1.5ex}{1.5ex}}}
\def\proof{\noindent{{\bf Proof }}}
\def\endproof{\hspace*{\fill}~\QED\par\endtrivlist\unskip}

\newtheorem{theorem}{Theorem}
\newtheorem{lemma}{Lemma}

\newtheorem{corollary}{Corollary}

\begin{document}

\maketitle
\thispagestyle{empty}
\pagestyle{empty}

\begin{abstract}

In this paper, we consider the standard discrete-time slotted ALOHA
with a finite number of terminals with infinite size buffers. In our
study, we jointly consider the stability of this system together
with the physical layer security. We conduct our studies on both
dominant and original systems, where in a dominant system each
terminal always has a packet in its buffer unlike in the original
system. For $N = 2$, we obtain the \textit{secrecy-stability}
regions for both dominant and original systems. Furthermore, we
obtain the transmission probabilities, which optimize system
throughput. Lastly, this paper proposes a new methodology in terms
of obtaining the joint stability and secrecy regions.

\end{abstract}

\section{INTRODUCTION}

Wireless multiple-access broadcast networks have received
significant interest from researchers in the past. Slotted ALOHA is
one of the basic class of such networks, and a large number of
random multiple access algorithms are devised as modifications of
this basic system. The two important issues of wireless systems,
i.e., the stability and security, have been separately studied in
the context of slotted ALOHA and the wireless broadcast networks.
Basically, stability requires that queue sizes remain finite when
time goes to infinity. Stability in slotted ALOHA has been
investigated in \cite{Abramson}, \cite{Bertsekas}. These results
have led to further studies, where various bounds and stability
regions are obtained for which the queues are stable
\cite{Tsybakov}, \cite{Rao}. In \cite{Malyshev}, sufficient and
necessary conditions for the stability of the system are obtained
and for two user case $(N=2)$, the stability region is identified.
More recent studies have aimed at obtaining tighter bounds for the
stability \cite{Luo}. On the other hand, secure communication over
physical layer was first introduced by Shannon \cite{Shannon} and
Wyner \cite{Wyner}. Recently, physically secure communication is
gaining more attention and a plethora of work have emerged on this
issue. The security of a single broadcast channel is investigated in
\cite{Csiszar}, \cite{Liu}. The secure rate allocation vectors are
determined for gaussian channels in \cite{TLiu}, \cite{Shafiee}.
\cite{YLiang} considers fading channels for which the perfect
secrecy regions and the optimal power allocation vectors maximizing
secrecy region, are obtained. These works investigated problems only
in the security context. In \cite{Liang}, stability and security are
combined in the context of wireless broadcast networks, where a base
station sends confidential messages to the users. In this system,
only downlink channels are considered, where no contention is taking
place.

In this paper, we jointly consider the stability and security issues
for slotted ALOHA systems, where each user wants to communicate with
a single base station. The broadcast channel is modeled as Rayleigh
fading channel, where $N$ user nodes send their confidential
messages to a base station as shown in Figure 1. Each message should
be kept secret from other users. Hence, all users except the
transmitting one are eavesdroppers. In the meantime, the stability
of the system should be maintained, i.e., the sizes of the queue for
each user should be finite when time goes to infinity. All
transmitters are assumed to be synchronized, and the reception of a
packet starts at the beginning of a slot and ends at the end of a
slot, i.e., each packet transmission occupies exactly one time slot.
Also, each user transmits with a probability, $q_i$, at each time
slot.
\begin{figure}[t]
\begin{center}
  \includegraphics[width=2.5in,height=2.2in]{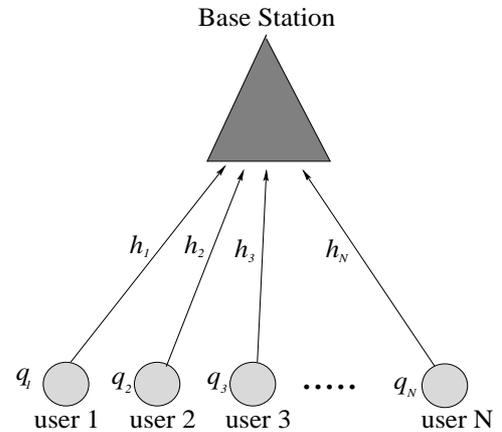}
\caption{Fading Broadcast Network}
\label{fig:1}       
\end{center}
\end{figure}

In prior studies, the secrecy region has always been studied at the
symbol level, and thus the secrecy region with respect to
transmission probabilities has not been examined before and this
perspective is introduced in our paper. Let us define
``secrecy-stability region" as the collection of transmission
probabilities, $q_i$, that satisfies both the stability and the
security conditions. Our goal is to find the secrecy-stability
region and obtain the optimal transmission probabilities, which
maximize the system throughput. For $N=2$, we specify the optimal
transmission probabilities for the \textit{dominant system}, where
it is assumed that users always have a packet to send in their
buffer. In the \textit{original system}; however the queues may not
always have a packet to transmit, and in this case, we show that the
maximized system throughput does not depend on the transmission
probabilities.

The rest of the paper is organized as follows: In Section II, we
introduce the channel model, and give the definitions of the secrecy
and the stability. In Section III and IV, we present our results for
the dominant and original systems. In Section V, we conclude the
paper by summarizing our contributions.

\section{Channel Model and Secrecy Capacity for Downlink Channels}

We consider a wireless broadcast network operating on a single
frequency channel. We assume that the channels from each user to
base station and other users are Rayleigh fading broadcast channels,
in which each output signals obtained by the base station and other
users are corrupted by multiplicative fading gains in addition to an
additive Gaussian noise as:
\begin{equation}
Y_{jn} = h_{ijn}X_i + w_{ijn}, \mbox{ for } 1 < i < N,
\label{channel model}
\end{equation}
where $X_i$ denotes the message transmitted by $i^{th}$ user,
$Y_{jn}$ is the channel output at user $j$, $h_{ijn}$ is the fading
coefficient for the channel between $i^{th}$ user and $j^{th}$ user,
and $w_{ijn}$ is Gaussian noise term with zero mean and unit
variance at the $n^{th}$ symbol time.

The secrecy level of confidential message, $W_i$, transmitted from
user $i$ to the base station is measured by the following {\em
equivocation rate} \cite{Liang}:
\begin{equation}
R_i \leq lim_{n \rightarrow \infty} \frac{1}{n} H(W_i |
Y_1^n,...,Y_{i-1}^n,Y_{i+1}^n,...,Y_N^n) \label{equivocation}
\end{equation}

The perfect secrecy is achieved when the transmission rate satisfies
(\ref{equivocation}) and {\em the secrecy region} is defined as the
set of all achievable rate vectors such that the perfect secrecy is
achieved \cite{Wyner}. In \cite{Liang}, the secrecy region of fading
broadcast channels in the downlink is obtained as:

\begin{eqnarray}
R_s = \begin{cases}&\bigcup(R_{1},R_{2},...,R_{N}): \\
& R_{i}\leq R_{s,i} = \underset{j\neq i, 1\leq j\leq N}{\min}
E_{\overline{h}\epsilon A(i)}
\begin{aligned}
&\left[\log(1+P|h_i|^2)\right. \\ 
&\left.-\log(1+P|h_{ij}|^2)\right]
\end{aligned},\\
& \mbox{for } 1 \leq i \leq N
\end{cases}
\label{secrecy region}
\end{eqnarray}
where $\overline{h}$ defines channel state, $h_i$ denotes the
channel gain between $i^{th}$ user and the base station, and
$h_{ij}$ denotes channel gain between $i^{th}$ and $j^{th}$ users.
$A(i)$ is the set of all channel states for which the channel gain
between $i^{th}$ user and the base station is the largest. In
addition, each user transmits with the same power level denoted as
$P$.

In addition, we define the stability of a queue as in \cite{Luo},
i.e., a queue is stable if it satisfies the following

\begin{eqnarray}
&&lim_{t\rightarrow\infty}Pr[s_i(t)<x]=F(x), \mbox{ and } \nonumber
\\
&& lim_{x\rightarrow\infty}F(x)=1,
\end{eqnarray}
where $s_i(t)$ is the size of the queue at time $t$. Namely, the
queue size should be finite at any time to achieve stability.

In the following, we investigate the secrecy-stability regions for
dominant and original systems.

\section{Dominant Aloha Uplink Channel}
The analysis of dominant systems was previously investigated in
\cite{Abramson}-\cite{Luo}. For the dominant system, it is shown
that the stability condition is as follows:

\begin{lemma} \cite{Tsybakov} If
\begin{equation}
\lambda_i < Q_i
\end{equation}
for all $i$ $(i=1,...,N)$, then the system is stable, where
$\lambda_i$ is the arrival rate and $Q_i$ is the successful
transmission probability for fading channels calculated as:
\begin{equation}
    Q_i=(1-p_{f,i})q_i \prod_{j=1\\ j\neq i}^N(1-q_j),
\end{equation}
where $p_{f,i}$ is the average failure probability of user $i$ due
to fading. Since $p_{f,i}$ is constant, we define $\lambda_i'$ as
$\lambda_i/p_{f,i}$. Then, the stability condition can be rewritten
as:
\begin{equation}
\lambda_i' < q_i \prod_{j=1\\ j\neq i}^N(1-q_j) \label{lemma_1}
\end{equation}
\end{lemma}

Lemma 1 specifies the stability condition, but does not give any
result about which specified arrival rates lead to a positive
stability region.

\begin{lemma} \cite{Tsybakov} The following condition should be satisfied to have positive
stability region:
    \begin{equation}
        x^{N-1}-\prod_{i=1}^N(x+\lambda_i')>0 \mbox{ for any } x>0
        \label{lemma 2}
    \end{equation}
For $N=2$, expression in (\ref{lemma 2}) can be written as:
\begin{equation}
        \sqrt{\lambda_1'}+\sqrt{\lambda_2'} < 1
        \label{lemma 2_2}
\end{equation}
\end{lemma}

The proofs of Lemma 1 and 2 can be found in \cite{Tsybakov}.

Now, we are ready to derive the perfect secrecy condition for
dominant ALOHA systems.

\begin{theorem} If
\begin{eqnarray}
\label{Theorem 1}
&& \rho_i \geq q_i \prod_{j=1, j\neq i}^N (1-q_j)  \\
&& \mbox{for } \rho_i = \frac{R_{s,i}}{R_i} \nonumber
\end{eqnarray}
for all $i$ $(i=1,...,N)$, then the system is secure. Note that
$\rho_i$ defines the ratio between the perfect secrecy capacity,
$R_{s,i}$, defined as in (\ref{secrecy region}) and the capacity of
fading channels, $R_i$.
\end{theorem}

\proof

The security region given in (\ref{secrecy region}) is computed for
downlink channels, where there is no channel contention. However, in
our model, we consider uplink channels, where packets when
transmitted simultaneously collide with each other. We assume that
collision results in scrambled bits and thus the received packets
cannot be correctly decoded. Therefore, we assume that the packets
in collision have no information value. Let us define new events
$Z_1^i$, $Z_2^i$ and $W_i$ as:

\begin{eqnarray}
Z_1^i &=&
\begin{cases}
1,  & \mbox{ transmission for } i^{th} \mbox{ user }\\
0,  & \mbox{ no transmission for } i^{th} \mbox{ user }
\end{cases}\nonumber \\
Z_2^i &=&
\begin{cases}
1,  & \mbox{ no collision for } i^{th} \mbox{ user }\\
0,  & \mbox{ collision for } i^{th} \mbox{ user }
\end{cases}\\
W_i &=&
\begin{cases}
\hat{W_i},  & \mbox{ if } Z_1^i = 1 \mbox{ and } Z_2^i = 1\\
0,       &  \mbox{ otherwise }\nonumber
\end{cases}
\end{eqnarray}
where $\hat{W_i}$ defines the event where messages are transmitted
with no collision.

The relationship between the equivocation rates of $W_i$ and
$\hat{W_i}$ is:

\begin{eqnarray}
\small &&H(\hat{W_i} | Y_1^n,...,Y_{i-1}^n,Y_{i+1}^n,...,Y_N^n)= \nonumber\\
&&P(Z_1^i=1,Z_2^i=1)H(W_i | Y_1^n,...,Y_{i-1}^n,Y_{i+1}^n,...,Y_N^n)
\label{branching}
\end{eqnarray}

Note that, the equivocation rate of $\hat{W_i}$ is the same as in
the downlink channels given in (\ref{secrecy region}).

We now determine a bound on transmission probabilities, $q_i$, as
follows:

\begin{eqnarray}
\small
R_i & \overset{(a)}{\leq}& \frac{H(W_i | Y_1^n,...,Y_{i-1}^n,Y_{i+1}^n,...,Y_N^n)}{n}  \nonumber \\
&\overset{(b)}{=}&\frac{H(\hat{W_i} | Y_1^n,...,Y_{i-1}^n,Y_{i+1}^n,...,Y_N^n)}{n} \frac{1}{P(Z_1^i=1,Z_2^i=1)}  \nonumber \\
&=& \frac{H(\hat{W_i} | Y_1^n,...,Y_{i-1}^n,Y_{i+1}^n,...,Y_N^n)}{P(Z_1^i=1)P(Z_2^i=1|Z_1^i=1)n} \nonumber\\
&\overset{(c)}{=}& \frac{1}{q_i\prod_{j\neq i}{(1-q_j)}}
\frac{H(\hat{W_i} | Y_1^n,...,Y_{i-1}^n,Y_{i+1}^n,...,Y_N^n)}{n}
 \nonumber \\
&\overset{(d)}{=}&\frac{1}{q_i\prod_{j\neq i}{(1-q_j)}} R_{s,i}
\label{uplink_secrecy}
\end{eqnarray}

where (a) follows from the perfect secrecy condition, (b) follows
from the branching property of entropy as in (\ref{branching}), (c)
is obtained by inserting the collision probability, (d) follows from
the fact that a successful transmission on the uplink has a secrecy
capacity equal to the secrecy capacity over a downlink channel.

\begin{figure}[!t]
\begin{center}
  \includegraphics[width=3.2in,height=2.0in]{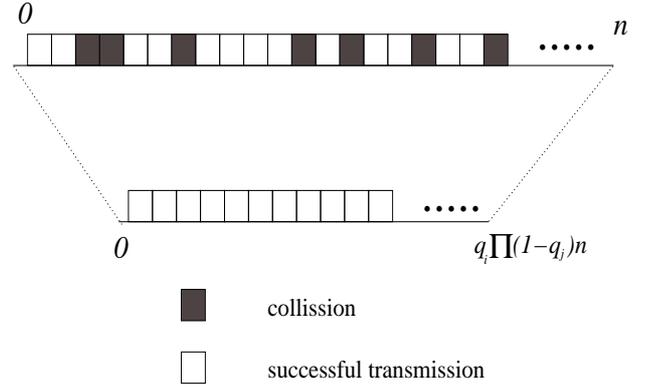}
\caption{Time Scaling for Secrecy Regions of Uplink and Downlink
Channels}
\label{fig:1}       
\end{center}
\end{figure}

Furthermore, in (\ref{uplink_secrecy}), $R_i$ is the capacity of
fading channels, which can be obtained by Shannon capacity,
$\log(1+P|h_i|^2)$ and $R_{s,i}$ is secrecy capacity of downlink
channels given by (\ref{secrecy region}).

By taking the ratio of secrecy capacity, $R_{s,i}$, and the shannon
capacity, $R_i$, we obtain the following perfect secrecy condition
in terms of the transmission probabilities:
\begin{equation}
\rho_i \geq q_i \prod_{j=1\\ j\neq i}^N(1-q_j) \label{secrecy
condition}
\end{equation}
\endproof

Intuitively, as shown in Figure 2, the condition in (\ref{secrecy
condition}) can be interpreted as the scaled version of a downlink
channel, when all slots with collisions are removed. In this case,
out of $n$ slots only $q_i\prod(1-q_j)$ portion of them carry
packets that can be correctly decoded, and the security condition
for this system is the same as the condition for downlink channel
with no contention.

After we obtain the secrecy and stability regions, the intersection
of these will give us the joint secrecy-stability region, where both
secrecy and stability are achieved.

\begin{corollary} If $\lambda_i'<\rho_i$, for all $i$ $(i=1,...,N)$, then there
exists positive secrecy-stability region.
\label{corollary}
\end{corollary}

Proof of this corollary can easily be derived from the secrecy and
the stability condition (\ref{Theorem 1}) and (\ref{lemma_1})
respectively.

Overall, Lemma 2 and Corollary 1 should be satisfied to have
positive secrecy-stability region.

For $N=2$, there are three different secrecy-stability regions as
illustrated in Figure 3, which lead to different solutions to the
maximization of system throughput. As seen in Figure 3(a), case 1
suggests more tighter secrecy bound and as a result we obtain
smaller secrecy-stability region. In this case, the channel with
eavesdropper is better compared to other cases. However, in case 3,
the secrecy condition loses its influence on the secrecy-stability
region, which is only equal to the stability region. In real life,
we may encounter with this case when there is a wall between
eavesdropper and transmitter, which worsens the channel.

\begin{figure}[!t]
\begin{center}
\begin{tabular}{c}
\includegraphics[width=1.5in,height=1.5in]{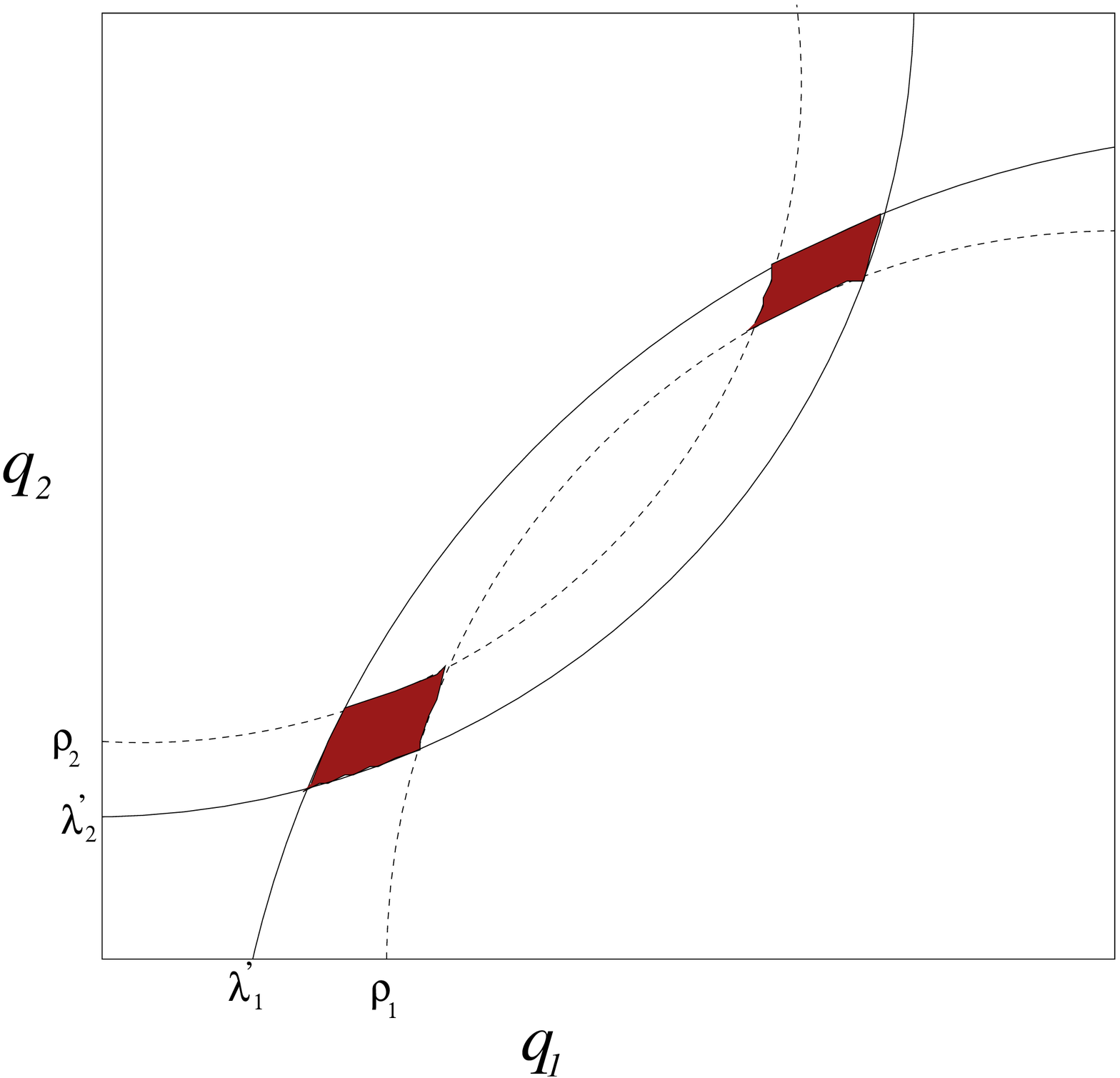} \\
(a) Case 1
\end{tabular}
\begin{tabular}{c}
\includegraphics[width=1.5in,height=1.5in]{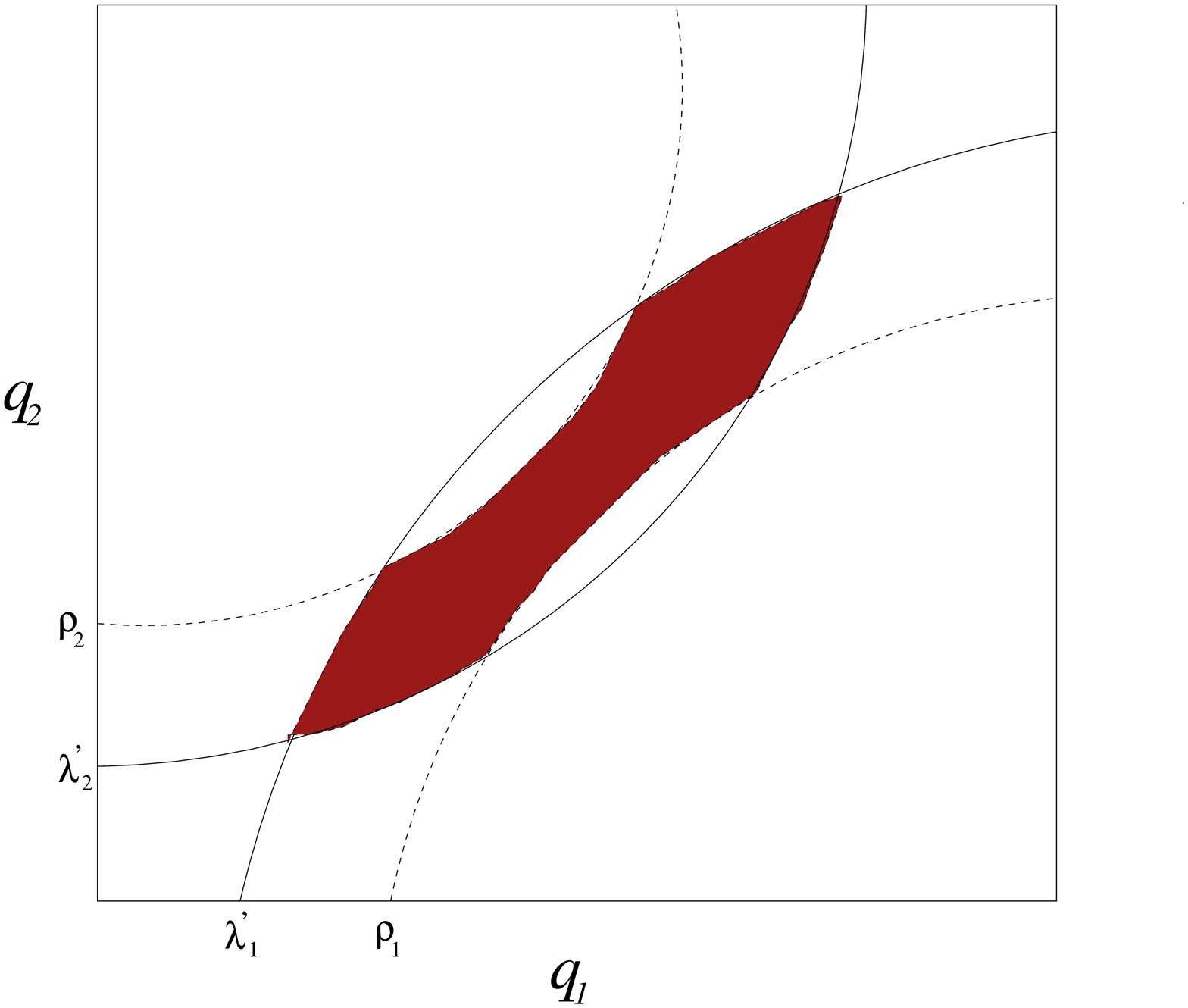}\\
(b) Case 2
\end{tabular}
\begin{tabular}{c}
\includegraphics[width=1.5in,height=1.5in]{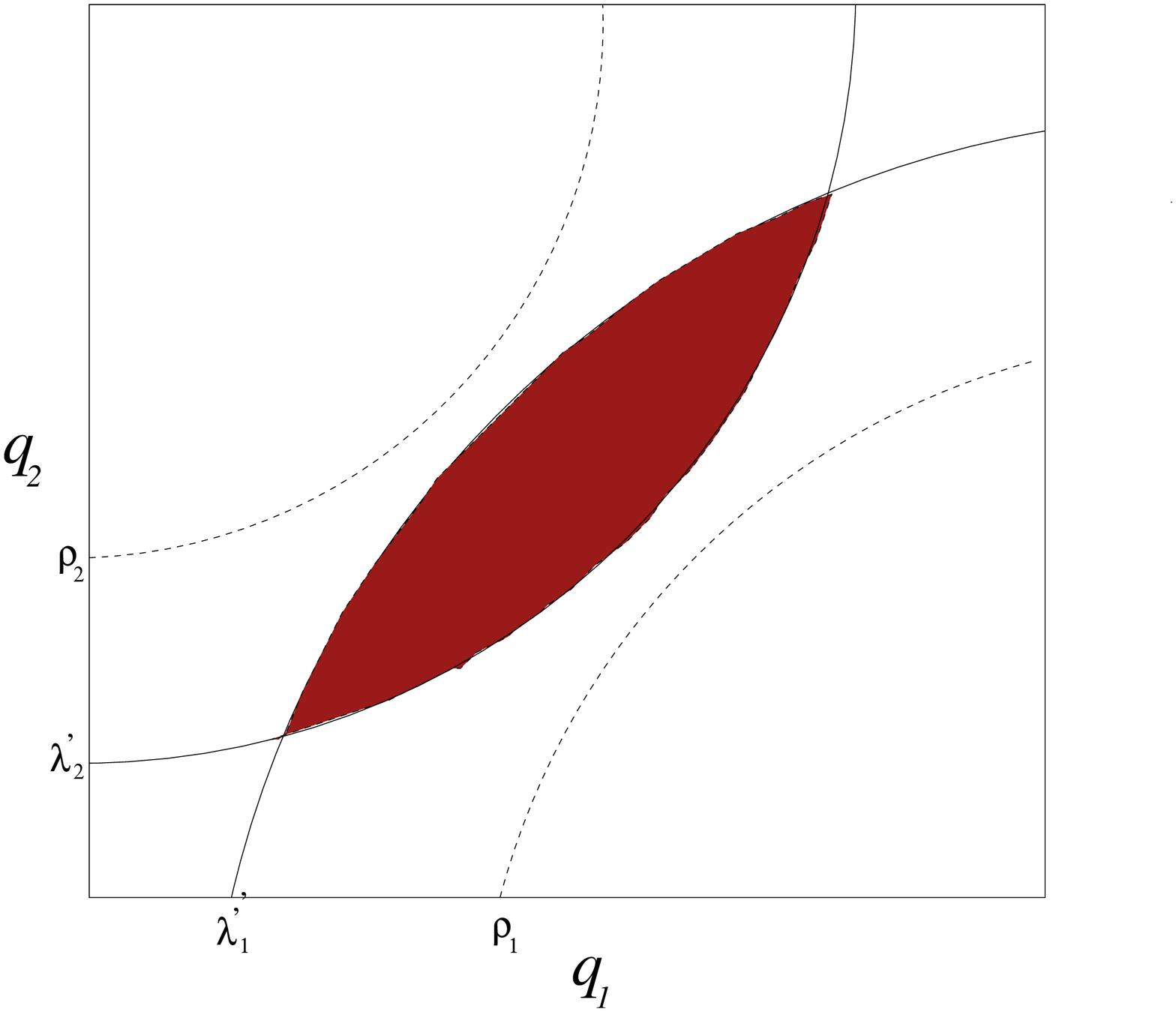} \\
(a) Case 3
\end{tabular}
\caption{Secrecy-Stability Regions for Dominant System}
\label{fig:3}
\end{center}
\end{figure}

\begin{theorem}

For $N=2$, the optimal transmission probabilities are as follows:

\textbf{(1)} when

\begin{equation}
        \sqrt{\rho_1}+\sqrt{\rho_2} \leq 1
\end{equation}

\begin{eqnarray}
q_1 =
\frac{(1+\rho_1)}{2}\bar{+}\frac{\sqrt{(1+\rho_1)^2-4(\rho_1+\rho_2)}}{2}
\nonumber \\
q_2 =
\frac{(1+\rho_2)}{2}\bar{+}\frac{\sqrt{(1+\rho_2)^2-4(\rho_1+\rho_2)}}{2}
\end{eqnarray}

\textbf{(2)} when
\begin{eqnarray}
        && \sqrt{\rho_1}+\sqrt{\rho_2} \geq 1 \nonumber \\
        && \sqrt{\rho_1}+\sqrt{\lambda_2'} < 1  \\
        && \sqrt{\rho_2}+\sqrt{\lambda_1'} < 1\nonumber
\end{eqnarray}

\begin{eqnarray}
        &&q_1=\sqrt{\rho_1}, \mbox{ } q_2 = 1- \sqrt{\rho_1} \mbox{ or }
        \nonumber \\
        &&q_1=1-\sqrt{\rho_2}, \mbox{ } q_2 = \sqrt{\rho_2}
\end{eqnarray}

\textbf{(3)} when
\begin{eqnarray}
        && \sqrt{\rho_1}+\sqrt{\rho_2} \geq 1 \nonumber \\
        && \sqrt{\rho_1}+\sqrt{\lambda_2'} > 1  \\
        && \sqrt{\rho_2}+\sqrt{\lambda_1'} > 1\nonumber
\end{eqnarray}

\begin{eqnarray}
        &&q_1=\sqrt{\lambda_1'}, \mbox{ } q_2 = 1- \sqrt{\lambda_1'} \mbox{ or }
        \nonumber \\
        &&q_1=1-\sqrt{\lambda_2'}, \mbox{ } q_2 = \sqrt{\lambda_2'}
\end{eqnarray}

\end{theorem}

\proof

The throughput optimization problem can be formulated as follows:

\begin{eqnarray}
    \label{objective_function}
    \mbox{max } S &=& q_1(1-q_2)(1-p_{f,1})  \\
    &&+q_2(1-q_1)(1-p_{f,2})  \nonumber\\
    \label{secrecy_con_user1}
    \mbox{s.t.} && q_1(1-q_2) \leq \rho_1\\
    \label{secrecy_con_user2}
    && q_2(1-q_1) \leq \rho_2  \\
    \label{stability_con_user1}
    && \lambda_1' < q_1(1-q_2) \\
    \label{stability_con_user2}
    && \lambda_2' < q_2(1-q_1) \\
    \label{optimization formulation}
    && 0 \leq q_1, q_2 \leq 1,
\end{eqnarray}
where  $p_{f,1}$  and $p_{f,2}$ denote the probability of channel
failure of the first and second users respectively.

The objective function in (\ref{objective_function}) can be
rewritten as a sum of two linear variables, e.g., ($X (\mbox{for
}q_1(1-q_2))+Y (\mbox{for }q_2(1-q_1))$), while these linear
variables are also constrained by linear inequalities. From the
basic knowledge of linear programming, the optimal solution is known
to be located at the corners of feasible region. Thus, as long as
$q_1$ and $q_2$ are in $[0,1]$, we expect that the optimal solution
is to appear on the boundary of the feasible region.

\textbf{(1)} First, we consider the case when the optimal solution
is achieved at the boundary of the secrecy region given in Figure
3(a). Then, the Lagrangian to solve optimization problem in
(\ref{objective_function})-(\ref{optimization formulation}) is given
by:
\begin{eqnarray}
L&=&q_1(1-q_2)(1-p_{f,1})+q_2(1-q_1)(1-p_{f,2}) \nonumber \\
&&-\beta_1(q_1(1-q_2)-\rho_1)-\beta_2(q_2(1-q_1)-\rho_2) \\
&&+\alpha_1(q_1(1-q_2)-\lambda_1')+\alpha_2(q_2(1-q_1)-\lambda_2')
\nonumber,
\end{eqnarray}
where $\beta_1$ and $\beta_2$ are lagrange multipliers for
inequalities in (\ref{secrecy_con_user1}) and
(\ref{secrecy_con_user2}), and $\alpha_1$ and $\alpha_2$ for
inequalities in (\ref{stability_con_user1}) and
(\ref{stability_con_user2}). Since the solution is assumed to be at
the boundary of secrecy region, $\alpha_1= 0$ and $\alpha_2 = 0$. We
take the derivative of the lagrangian with respect to non-zero
lagrange multipliers and transmission probabilities, and equate to
zero as:
\begin{eqnarray}
\frac{\partial L}{\partial q_1} &=& (1-q_2)(1-p_{f,1})-q_2(1-p_{f,2})\nonumber\\
&&-\beta_1(1-q_2)+\beta_2q_2 = 0 \nonumber\\
\frac{\partial L}{\partial q_2} &=& (1-q_1)(1-p_{f,2})-q_1(1-p_{f,1})\nonumber\\
&&-\beta_2(1-q_1)+\beta_1q_1 = 0 \nonumber \\
\frac{\partial L}{\partial \beta_1} &=& q_1(1-q_2)-\rho_1 = 0 \nonumber \\
\frac{\partial L}{\partial \beta_2} &=& q_2(1-q_1)-\rho_2 = 0
\end{eqnarray}

By simple manipulations, we obtain $\beta_1$ and $\beta_2$ as 1,
which satisfies the condition that lagrange multipliers should be
greater than zero. For this case, we found $q_1$ as
$\frac{(1+\rho_1)}{2}\bar{+}\frac{\sqrt{(1+\rho_1)^2-4(\rho_1+\rho_2)}}{2}$
and $q_2$ as
$\frac{(1+\rho_2)}{2}\bar{+}\frac{\sqrt{(1+\rho_2)^2-4(\rho_1+\rho_2)}}{2}$.
Note that this solution attains a real root, when the following
conditions are satisfied : $(1+\rho_1)^2-4(\rho_1+\rho_2) \geq 0$
and $(1+\rho_2)^2-4(\rho_1+\rho_2) \geq 0$. After some
manipulations, we see that a real solution is realized when

\begin{equation}
\sqrt{\rho_1}+\sqrt{\rho_2} \leq 1 \label{condition_case1}
\end{equation}

\textbf{(2)} Figure 3(b) shows the secrecy-stability region, when
the condition in (\ref{condition_case1}) does not hold, i.e.,
$\beta_1
>0$ and $\beta_2>0$ jointly cannot be satisfied.

First, Let $\beta_1 \geq 0$ and $\beta_2 = 0$, then we have the
following derivatives:
\begin{eqnarray}
\frac{\partial L}{\partial q_1} &=& (1-q_2)(1-p_{f,1})-q_2(1-p_{f,2})-\beta_1(1-q_2) = 0 \nonumber\\
\frac{\partial L}{\partial q_2} &=& (1-q_1)(1-p_{f,2})-q_1(1-p_{f,1})+\beta_1q_1 = 0 \nonumber \\
\frac{\partial L}{\partial \beta_1} &=& q_1(1-q_2)-\rho_1 = 0
\label{case2 derivatives}
\end{eqnarray}

From the first two equations in (\ref{case2 derivatives}), we find
$q_1 = 1-q_2$ and by using the third equation in (\ref{case2
derivatives}), we obtain the solution as: $q_1 = \sqrt{\rho_1}$ and
$q_2 = 1- \sqrt{\rho_1}$. However, this solution should satisfy the
stability condition in (\ref{stability_con_user2}) as well:
\begin{eqnarray}
    q_2(1-q_1) &>& \lambda_2' \nonumber \\
    (1- \sqrt{\rho_1})(1- \sqrt{\rho_1}) &>& \lambda_2' \nonumber \\
    \sqrt{\rho_1} + \sqrt{\lambda_2'} &<& 1
    \label{condition_case2}
\end{eqnarray}

Similarly, when $\beta_2 \geq 0$ and $\beta_1 = 0$, we can follow
the same discussion as before to obtain the solution as: $q_2 =
\sqrt{\rho_2}$ and $q_1 = 1-\sqrt{\rho_2}$ when $ \sqrt{\rho_2} +
\sqrt{\lambda_1'} < 1 $.

Now, the optimal solution is one of these two solutions; however,
since the secrecy-stability region is not convex, we cannot
determine the optimal closed form solution.

\textbf{(3)} When the conditions in (\ref{condition_case1}) and
(\ref{condition_case2}) do not hold, the secrecy-stability region
only consists of the stability region as shown in Figure 3(c). Then,
we have the following optimization problem:

\begin{eqnarray}
    \mbox{max } && S = q_1(1-q_2)(1-p_{f,1})+q_2(1-q_1)(1-p_{f,2}) \nonumber \\
    \mbox{s.t.} && \lambda_1' < q_1(1-q_2) \nonumber \\
         && \lambda_1' < q_2(1-q_1)  \nonumber \\
         && 0 \leq q_1, q_2 \leq 1,
\end{eqnarray}

 As before, we expect that the optimal solution is to appear at the boundary. Both
constraints cannot be active, so we select only one of them as
active. First, we consider the first constraint as the active
constraint: The lagrange multipliers are: $\alpha_1 \leq 0 $ and
$\alpha_2 = 0$. Then, the lagrange function is as follows:
\begin{eqnarray}
 L &=& q_1(1-q_2)(1-p_{f,1})+q_2(1-q_1)(1-p_{f,2}) \\
 &&+ \alpha_1(q_1(1-q_2)-\lambda_1') \nonumber
\end{eqnarray}
Then, we have the following derivatives:

\begin{eqnarray}
\frac{\partial L}{\partial q_1} &=& (1-q_2)(1-p_{f,1})-q_2(1-p_{f,2})+\alpha_1(1-q_2)\nonumber = 0 \\
\frac{\partial L}{\partial q_2} &=& (1-q_1)(1-p_{f,2})-q_1(1-p_{f,1})-\alpha_1q_1\nonumber = 0 \\
\frac{\partial L}{\partial \alpha_1} &=& q_1(1-q_2)-\lambda_1' = 0 \nonumber \\
\end{eqnarray}

If we solve these equations, we obtain the solution as: $q_1 =
\sqrt{\lambda_1'}$ and $q_2 = 1-\sqrt{\lambda_1'}$.

Similarly, if we let $\alpha_2 \leq 0 $ and $\alpha_1 = 0$, then we
get the solution as follows: $q_1 = 1- \sqrt{\lambda_2'}$ and $q_2 =
\sqrt{\lambda_2'}$.

\endproof

\section{Original Aloha Uplink Channel}
In this section, we consider systems where the buffers of the users
do not always have packets. Let $p_{e,i}$ be the probability of
queue of user $i$ being empty. Then, the secrecy condition is
defined as follows:

\begin{theorem} If

\begin{equation}
    q_i\prod_{j=1, j\neq i}^N((1-p_{e,j})(1-q_j)+p_{e,j}) \leq \rho_i
\end{equation}
for all $i$ $(i=1,...,N)$, then the system is secure.
\end{theorem}
\proof

In Theorem 1, we obtained the secrecy condition for dominant
systems, where there are no empty queues. The method of the proof of
Theorem 3 is the same, where we want to determine the portion of
time when there is a single transmission and no collision. However,
in the original system the probability of this event is different
from the one in a dominant system. Let us define new event, $E_i$
as:

\begin{eqnarray}
E_i =
\begin{cases}
1,  & \mbox{ queue of } i^{th} \mbox{ user is not empty}\\
0,  & \mbox{ queue of } i^{th} \mbox{ user is empty}
\end{cases}
\end{eqnarray}

Then, the equivocation rate for the original system is

\begin{eqnarray}
\small &&H(\hat{W_i} | Y_1^n,...,Y_{i-1}^n,Y_{i+1}^n,...,Y_N^n)=
 \\
&&P(Z_1^i=1,Z_2^i=1|E_i=1)H(W_i |
Y_1^n,...,Y_{i-1}^n,Y_{i+1}^n,...,Y_N^n), \nonumber
\label{branching_2}
\end{eqnarray}
where
\begin{eqnarray}
\scriptsize
    &&P(Z_1^i=1,Z_2^i=1|E_i=1) = \nonumber \\
    && q_i\prod_{j=1, j\neq i}^N((1-p_{e,j})(1-q_j)+p_{e,j})
\end{eqnarray}

If we make the same mathematical operations as in
(\ref{uplink_secrecy}), we obtain the following secrecy condition:
\begin{eqnarray}
    q_i\prod_{j=1, j\neq
    i}^N((1-p_{e,j})(1-q_j)+p_{e,j}) \leq \rho_i
\end{eqnarray}

\endproof
Note that $\rho_i$ can be interpreted as proportion of time in all
occupied slots with successful transmissions with no collisions.

\begin{theorem} For $N=2$, the secrecy condition is as follows:

\begin{eqnarray}
    && \frac{1+\lambda_1'-\lambda_2'-\sqrt{\rho_1^2((\lambda_2'-1-\lambda_1')^2-4\lambda_1')}}{2\lambda_1'}
    = q_1^* \leq q_1 \nonumber \\
   && \mbox{ for } q_2
   \geq\frac{\lambda_2'\rho_1}{\rho_1-q_1^*\lambda_1'}=q_2^{**}
   \label{Theorem 4}
\end{eqnarray}

Due to the symmetric behavior of the system, the secrecy condition
for the second user is obtained by replacing $\rho_1$ by $\rho_2$,
$\lambda_1'$ by $\lambda_2'$ and $\lambda_2'$ by $\lambda_1'$ in
(\ref{Theorem 4}).
\end{theorem}

\proof

In Theorem 3, we have shown that the system is secure when

\begin{equation}
    q_1 \leq \frac{\rho_1}{(1-p_{e,2})(1-q_2)+p_{e,2}}
    \label{Theorem2_again}
\end{equation}

Also by Little's theorem \cite{Bertsekas}, we know that

\begin{equation}
    p_{e,1}=1-\frac{\lambda_1}{\mu_1},
    \label{empty_queue}
\end{equation}
where $\mu_1$ is the average service rate of the first user. We have
the following relationship between the service rate, $\mu_1$, and
$\rho_1$:
\begin{equation}
    \mu_1 = (1-p_{f,1})q_1((1-p_{e,2})(1-q_2)+p_{e,2}) \leq \rho_1(1-p_{f,1})
    \label{service_rate_rho}
\end{equation}

Thus, by substituting (\ref{service_rate_rho}) into
(\ref{empty_queue}), we obtain
\begin{equation}
    p_{e,1} \leq 1-\frac{\lambda_1}{\rho_1(1-p_{f,1})}=
    1-\frac{\lambda_1'}{\rho_1}
\end{equation}

By the symmetric behavior of the system, we know that
\begin{eqnarray}
     p_{e,2} &=& 1-\frac{\lambda_2'}{q_2((1-p_{e,1})(1-q_1)+p_{e,1})}
     \nonumber \\
             &\geq& 1-\frac{\lambda_2'}{q_2(\frac{\lambda_1'}{\rho_1}(1-q_1)+1-\frac{\lambda_1'}{\rho_1})}
             \label{empty_queue2}
\end{eqnarray}

By substituting (\ref{empty_queue2}) into (\ref{Theorem2_again}), we
obtain the following quadratic equation:
\begin{equation}
    \lambda_1' q_1^2+\rho_1(\lambda_2'-1-\lambda_1')q_1+\rho_1^2 \leq 0
    \label{quadratic_eqn}
\end{equation}

Interestingly, the above equation does not depend on the
transmission probability of eavesdropper, $q_2$. From
(\ref{quadratic_eqn}), we obtain a bound on $q_1$ as:
\begin{eqnarray}
\footnotesize
   &&
   \mbox{max}(0,\frac{1+\lambda_1'-\lambda_2'-\sqrt{\rho_1^2((\lambda_2'-1-\lambda_1')^2-4\lambda_1')}}{2\lambda_1'}) \leq q_1 \nonumber \\
    && \leq \mbox{min}(1,\frac{1+\lambda_1'-\lambda_2'+\sqrt{\rho_1^2((\lambda_2'-1-\lambda_1')^2-4\lambda_1')}}{2\lambda_1'})
    \label{solution_quadratic}
\end{eqnarray}

Also note that, the term,
$\frac{1+\lambda_1'-\lambda_2'-\sqrt{\rho_1^2((\lambda_2'-1-\lambda_1')^2-4\lambda_1')}}{2\lambda_1'}$,
is positive, since $\rho_1 \leq 1 $ and $\lambda_1' \geq 0$. In
addition, the term,
$\frac{1+\lambda_1'-\lambda_2'+\sqrt{\rho_1^2((\lambda_2'-1-\lambda_1')^2-4\lambda_1')}}{2\lambda_1'}$,
is always bigger than one, since from lemma 2 we know that
$\sqrt{\lambda_1'}+\sqrt{\lambda_2'} < 1$ and so
$\lambda_1'+\lambda_2' < 1$. Then, the solution in
(\ref{solution_quadratic}) becomes:

\begin{equation}
  \frac{1+\lambda_1'-\lambda_2'-\sqrt{\rho_1^2((\lambda_2'-1-\lambda_1')^2-4\lambda_1')}}{2\lambda_1'})
    = q_1^* \leq q_1
\end{equation}

Also, $p_{e,2} \geq 0$ which results in (\ref{empty_condition}) by
substituting $q_1^*$ in (\ref{empty_queue2})
\begin{equation}
    q_2\geq\frac{\lambda_2'\rho_1}{\rho_1-q_1^*\lambda_1'}=q_2^{**},
    \label{empty_condition}
\end{equation}

Note that, when $q_1$ is equal to $q_1^*$ and $q_2$ is $q_2^{**}$,
then $p_{e,2}$ is zero, which means that the second user always has
a packet to transmit as in a dominant system.

Finally, we attain the following condition:

\begin{eqnarray}
    &&\frac{1+\lambda_1'-\lambda_2'-\sqrt{\rho_1^2((\lambda_2'-1-\lambda_1')^2-4\lambda_1')}}{2\lambda_1'})
    = q_1^* \leq q_1 \nonumber \\
   && \mbox{ for } q_2 \geq\frac{\lambda_2'\rho_1}{\rho_1-q_1^*\lambda_1'}
\end{eqnarray}

\endproof
\begin{lemma} In order to have a stable system, the average service rate,
$\mu_i$, should be greater than the arrival rate. Then, we have the
following stability condition:
\begin{eqnarray}
    &&\mu_i = (1-p_{f,i})q_i \prod_{j=1, j\neq i}^N((1-p_{e,j})(1-q_j)+p_{e,j})
    >
    \lambda_i \nonumber\\
    &&q_i \prod_{j=1, j\neq i}^N((1-p_{e,j})(1-q_j)+p_{e,j}) >
        \frac{\lambda_i}{1-p_{f,i}} = \lambda_i'
\end{eqnarray}
\end{lemma}

The secrecy and stability regions for the original system are shown
in Figure 4(a) and Figure 4(b) respectively. The proof for the
stability region can be found in \cite{Tsybakov}.

By combining both regions, we obtain the secrecy-stability region as
illustrated in Figure 4(c). At the points, $(q_1^{*}, q_2^{**})$ and
$(q_1^{**}, q_2^{*})$, the probability of queue 2 and queue 1 being
empty are zero, which is the same in the dominant system. Thus, the
point $(q_1^{**}, q_2^{**})$  is located on the intersection of two
stability curves as seen in Figure 4(c).


\begin{figure}[!t]
\begin{center}
\begin{tabular}{c}
\includegraphics[width=1.5in,height=1.5in]{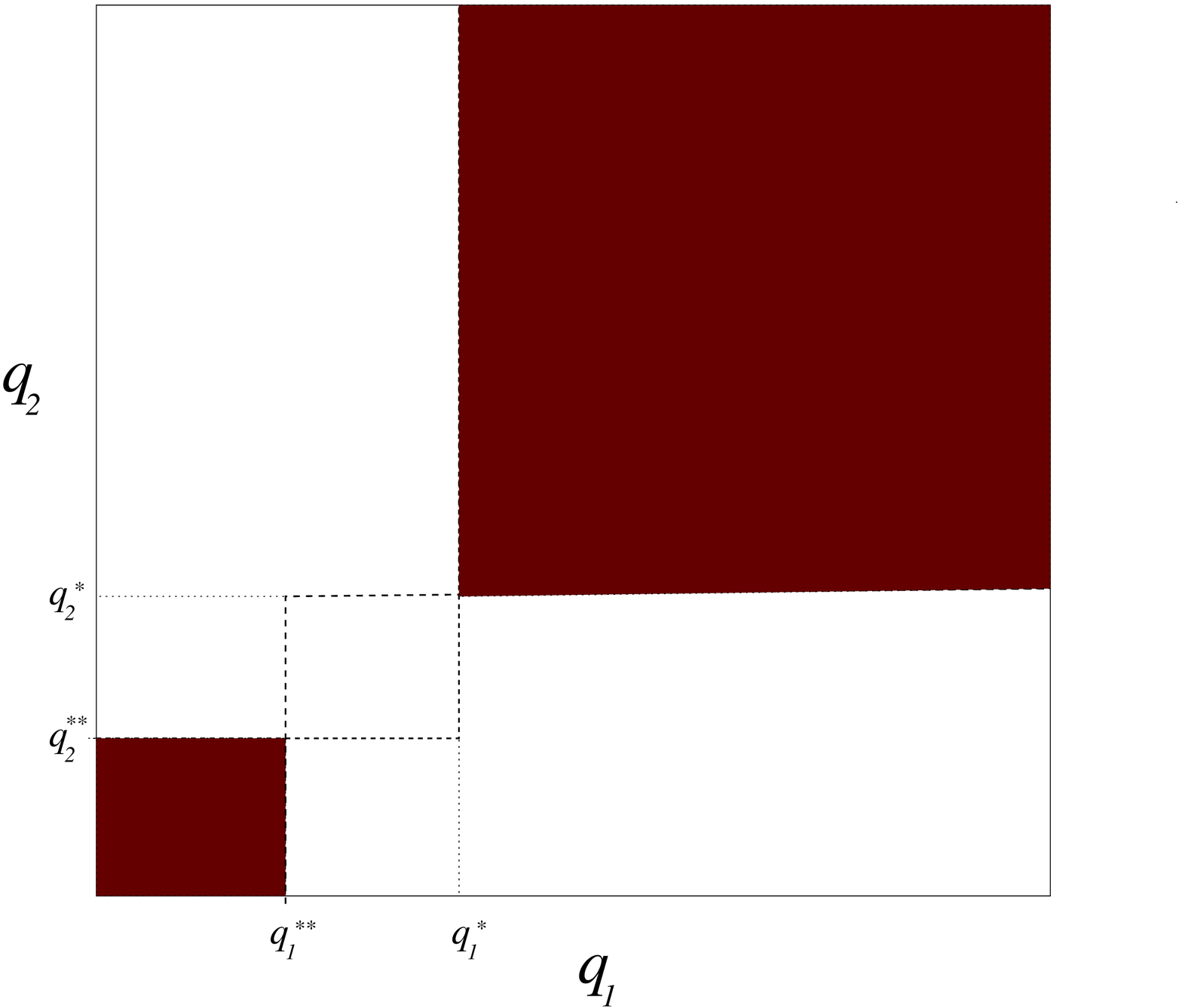} \\
(a) Secrecy Region
\end{tabular}
\begin{tabular}{c}
\includegraphics[width=1.5in,height=1.5in]{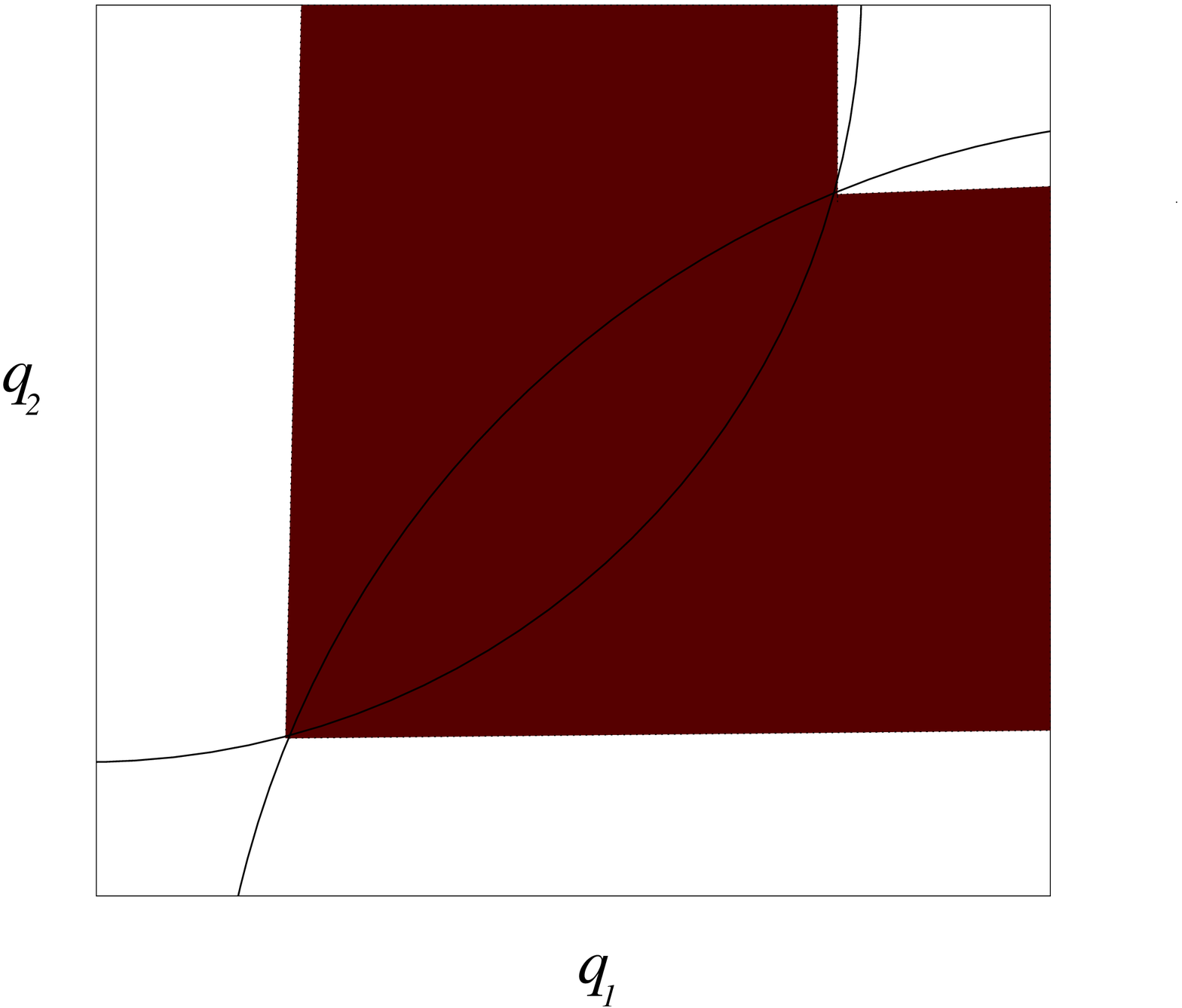}\\
(b) Stability Region
\end{tabular}
\begin{tabular}{c}
\includegraphics[width=1.5in,height=1.5in]{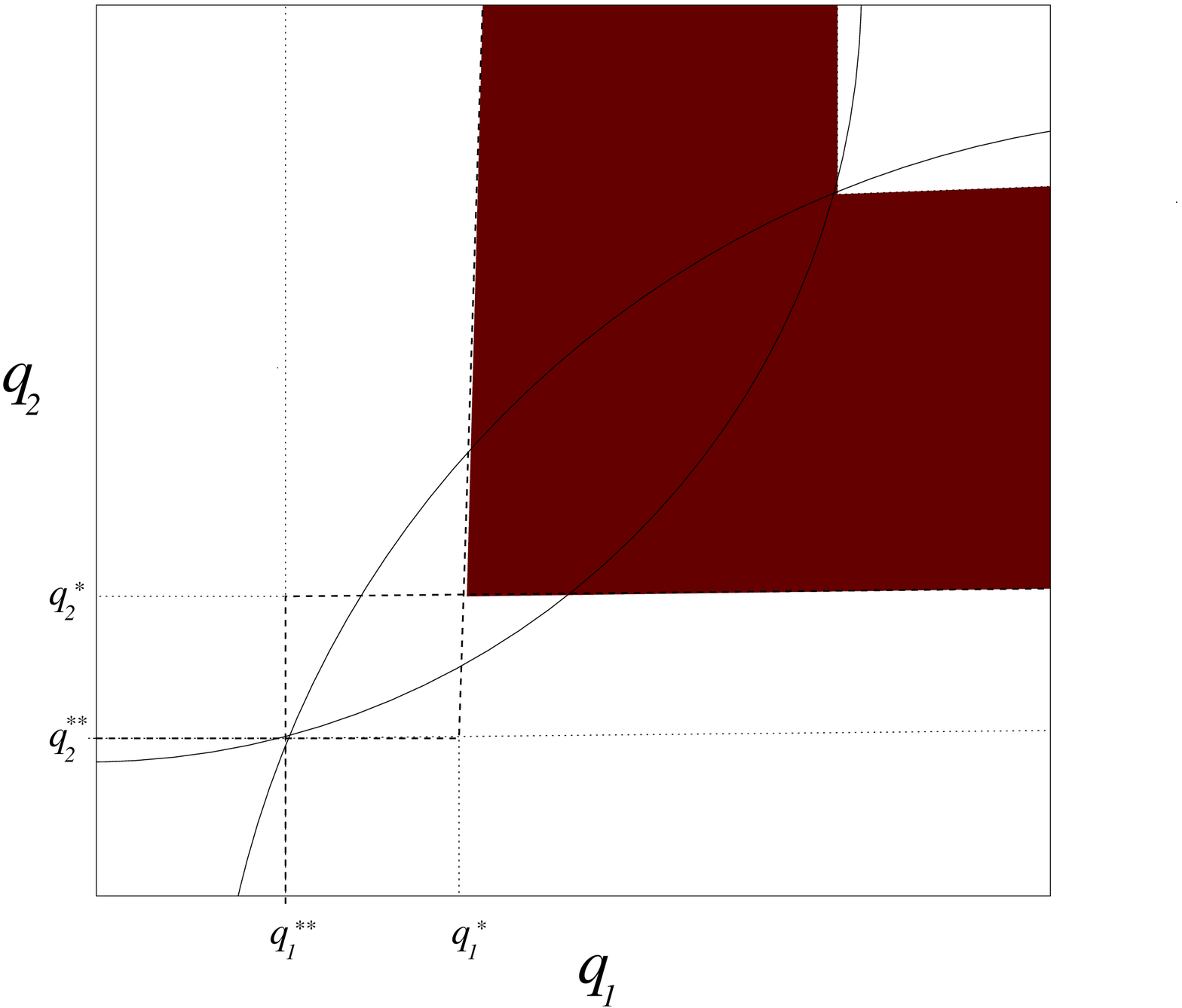} \\
(c) Secrecy-Stability Region
\end{tabular}
\caption{Secrecy and Stability Regions for Original System}
\label{fig:3}
\end{center}
\end{figure}

\begin{theorem} The optimum throughput, $S^*$, for any transmission
probabilities in the secrecy-stability region is equal to sum of
arrival rates:
\begin{equation}
    S^* = \sum_{i=1}^N \lambda_i
\end{equation}

\end{theorem}

\proof

The system throughput is formulated as:

\begin{equation}
    S = \sum_{i=1}^N(1-p_{f,i})(1-p_{e,i})q_i \prod_{j\neq
    i}\left[(1-p_{e,j})(1-q_j)+p_{e,j}\right]
\end{equation}
We know that $1-p_{e,i}$ is equal to $\lambda_i/ \mu_i$ and the
term, $(1-p_{f,i})q_i\prod_{j\neq i}(1-p_{e,j})(1-q_j)+p_{e,j}$, is
defined as  the average service rate, then for $\mu_i > 0$ we obtain
the following result:

\begin{eqnarray}
    S &=& \sum_{i=1}^N\frac{\lambda_i}{\mu_i}\mu_i
    \nonumber \\
      &=& \sum_{i=1}^N \lambda_i
\end{eqnarray}

\endproof

In theorem 5, we find out that the optimum throughput  can be any
point in the secrecy-stability region. That is because, increasing
the transmission probabilities leads to a decrease in the
probability of having empty queue, and this results in a decrease in
successful transmission probability. Thus, even if we increase the
transmission opportunities, success out of these opportunities will
not change.

\section{Conclusion}

In this paper, we have studied slotted ALOHA network, for which we
have obtained secrecy-stability conditions for the dominant and
original system. We have further obtained the optimal transmission
probabilities for $N=2$. This is the first work that jointly
addresses both the secrecy and stability of a wireless network with
contention.

\end{document}